\documentclass[graybox,natbib,nosecnum]{svmult}
\bibpunct{(}{)}{;}{a}{}{,} 

\pdfoutput=1   

\usepackage{mathptmx}       
\usepackage{helvet}         
\usepackage{courier}        
\usepackage{type1cm}        

\usepackage{makeidx}         
\usepackage{graphicx}        
\usepackage{multicol}        
\usepackage[bottom]{footmisc}
\usepackage[normalem]{ulem}	
\usepackage{hyperref}  
\usepackage{color}  

\usepackage{soul}   

\newcommand*\rd{}
\newcommand*\aap{A\&A}

\newcommand*\aj{AJ}

\newcommand*\apj{ApJ}
\newcommand*\apjl{ApJ}

\newcommand*\araa{ARA\&A}

\newcommand*\mnras{MNRAS}

\newcommand*\nat{Nature}

\newcommand*\procspie{Proc SPIE}

\def\note #1]{{\bf #1]}}

\def\msun{\,{\rm M}_\odot}

\makeindex             

\begin{document}

\title*{Ages for exoplanet host stars}
\author{J{\o}rgen Christensen-Dalsgaard and V\'ictor Silva Aguirre}
\institute{J{\o}rgen Christensen Dalsgaard \at Stellar Astrophysics Centre, Department of Physics and Astronomy, Aarhus University, Ny Munkegade 120, 8000 Aarhus C, Denmark, \email{jcd@phys.au.dk} \and Victor Silva Aguirre \at Stellar Astrophysics Centre, Department of Physics and Astronomy, Aarhus University, Ny Munkegade 120, 8000 Aarhus C, Denmark, \email{victor@phys.au.dk}}
%
%
\maketitle

\abstract{
Age is an important characteristic of a planetary system, but also one that is difficult to determine. Assuming that the host star and the planets are formed at the same time, the challenge is to determine the stellar age. Asteroseismology provides precise age determination, but in many cases the required detailed pulsation observations are not available. Here we concentrate on other techniques, which may have broader applicability but also serious limitations. Further development of this area requires improvements in our understanding of the evolution of stars and their age-dependent characteristics, combined with observations that allow reliable calibration of the various techniques.
}

\section{Introduction }
%
%
The age is an obviously important part of the characterization of an exo\-planetary system. It determines the time available for the system's evolution, including the processes that have resulted in its present configuration; an example is the analysis by \citet{triaud_11} of the age dependence of orbital inclination, measured with the Rossiter-McLaughlin effect. Also, the age of the system sets the scene for the evolution of life on possible habitable planets. Determination of such ages must in almost all cases be based on inference of the age of the central star of the system. The so far only known exception is the solar system, where precise, and probably accurate, ages can be obtained from analysis of radioactive isotopes, even detailing the ages of individual components of the 
original planetary disk \citep[e.g.,][]{connelly_etal_12}. In contrast, age determination for distant stars is always indirect, depending on the observation of some property that changes with time, such that the determination will in the end always depend on stellar modelling and the associated potential systematic uncertainties. For stars in and just after the phase of central hydrogen burning, the so-called main sequence, the change in stellar properties is based on the gradual change in the composition of the stellar core as hydrogen fuses into helium. However, there are secondary changes in stellar properties, particularly related to the evolution of rotation and magnetic activity, which also provide potential age diagnostics. {\rd Extensive reviews of stellar age determination were provided by \citet{soderblom_10} and \citet{jeffries_14}}, while \citet{soderblom_etal_14} focussed on the challenges of ages of young stars.

As discussed in the chapter {\it Characterizing host stars using asteroseismology} (see also below) asteroseismic analysis of observations of stellar oscillations provides a relatively precise measure of stellar age which is directly related to stellar internal properties and hence to the nuclear evolution of the star. Space-based photometry with CoRoT and, in particular, {\it Kepler}, has been an excellent source of oscillation observations, leading to asteroseismic characterization, including determination of ages, for a number of exo\-planet-host stars \citep[e.g.,][]{silva_aguirre_etal_15}. However, such observations of main-sequence stars require a rapid cadence and good signal-to-noise ratio, to detect the faint oscillation signal; also, stars cooler than about spectral type K generally do not show oscillations. As a result, asteroseismic age determinations have been limited to the brightest and best-observed {\it Kepler} targets. Ages of fainter or cooler stars must be determined with other methods, possibly calibrated with asteroseismic data. These methods are the focus of the present chapter.
\section{Observational properties of stars}
Here we focus on `classical' properties obtained from photometric and spectroscopic observations, and their relation to the intrinsic  stellar properties. For a detailed overview of astrophysical observations, see, for example, \citet{chromey_16}. From a global perspective stellar models are characterized by the mass $M$, the surface radius $R$ and the surface luminosity $L$. In addition, the temperature in the stellar atmosphere is represented by {\it the effective temperature} $T_{\rm eff}$ defined, according to the relation for a black body, by
\begin{equation}
L = 4 \pi R^2 \sigma T_{\rm eff}^4 \; ,
\label{eq:teff}
\end{equation}
where $\sigma$ is the Stefan-Boltzmann constant. An important quantity is also the surface gravity
\begin{equation}
g = {G M \over R^2} \; ,
\label{eq:surfgrav}
\end{equation}
where $G$ is the gravitational constant; when characterizing observations $g$ is most often represented by $\log g$, with $\log$ being logarithm to base 10. Finally, the composition is typically specified by the abundances $X, Y$ and $Z$ by mass of hydrogen, helium and elements heavier than helium, often called `metals'; these quantities satisfy the obvious relation $X + Y + Z = 1$.

Photometric observations are typically characterized in terms of {\it magnitudes} $m$, related to the observed flux $f$ by $m = -2.5 \log f + c$, where $c$ is a constant which in practice is determined from suitable standard stars. Here the observed flux depends on the properties of the instrument, including the choice of filters, the absorption in the atmosphere, and the intrinsic distribution of energy with wavelength in the stellar spectrum. Different magnitude scales are defined by different choices of filters, collected in systems designed to provide a reasonable characterization of the energy distribution. An example is the Johnson system with magnitudes denoted $U, B$ and $V$ sensitive to the ultraviolet, blue and visual parts of the spectrum, respectively. Since the energy distribution roughly follows a black-body (Planck) curve, differences between magnitudes in different bands, or {\it colour indices} such as $B - V$, provide a measure of $T_{\rm eff}$. Also, from a properly calibrated determination of the magnitude, together with a bolometric correction relating this to the total energy received at Earth from the star and a determination from parallax measurement of its distance, the stellar total luminosity $L$ can be determined. Stellar distance determinations will be greatly improved by the very accurate parallax observations from the Gaia mission \citep[see e.g.,][]{GaiaCollaboration_2016}.

Spectroscopic observations determine the absorption-line profiles which provide a measure of the surface composition of the star. In cooler stars, typical of exo\-planet hosts, lines of helium are not visible. Hence the composition is characterized by the ratio $Z_{\rm s}/X_{\rm s}$ between the abundance of heavy elements and hydrogen. Also, observations are often measured relative to the Sun, motivating the definition of {\it the metallicity} as
\begin{equation}
[{\rm M/H}] = \log \left( {Z_{\rm s}/X_{\rm s} \over
Z_{\rm s,\odot}/X_{\rm s,\odot}} \right) \; .
\end{equation}
Since the iron abundance is typically well characterized by the observations it is common to replace $[{\rm M/H}]$ by $[{\rm Fe/H}]$. It should be noted, however, that the transition from the observed $[{\rm Fe/H}]$ to $Z_{\rm s}/X_{\rm s}$ requires knowledge of $Z_{\rm s,\odot}/X_{\rm s,\odot}$, which has undergone recent revisions \citep[e.g.,][]{asplund_etal_09} leading to serious problems with solar modelling \citep[for a review, see][]{basu_antia_08}. An additional uncertainty is the helium abundance; this is often obtained, although with no strong justification, by relating it to $Z$ through an assumed ratio $\Delta Y/Z$ between the increase in $Y$ and $Z$ since big-bang nucleosynthesis, obtained from Galactic chemical evolution. The relative strengths of the spectral lines also depend on the ionization and excitation states of the atoms in the stellar atmosphere and hence provide a measure of $T_{\rm eff}$, while the line broadening depends on the density and hence gravity in the stellar atmosphere, such that the line widths can be used to determine $\log g$. The detailed analyses to determine $T_{\rm eff}$, $\log g$ and composition utilize fits to computed stellar atmosphere models. In the following we generally neglect the complications of the photometric and spectroscopic analyses and assume that the stars can be characterized by $T_{\rm eff}$, $L$ (or $\log\,g$), and their surface metallicity.

We finally note that stellar radii can be obtained from interferometric observations \citep[e.g.,][and references therein]{Hanbury_Brown_74,Segransan_2003,Boyajian_2012}, combined with determination of the distance of the star. This will only be available for a fairly limited sample of stars but plays an important role in the calibration of other techniques to characterize stellar properties \citep[see e.g.,][]{Huber_2012,White_2013,Casagrande_2014}.
%
%
\section{Ages from stellar evolution}
\begin{figure}
\centering
\includegraphics[width=\textwidth]{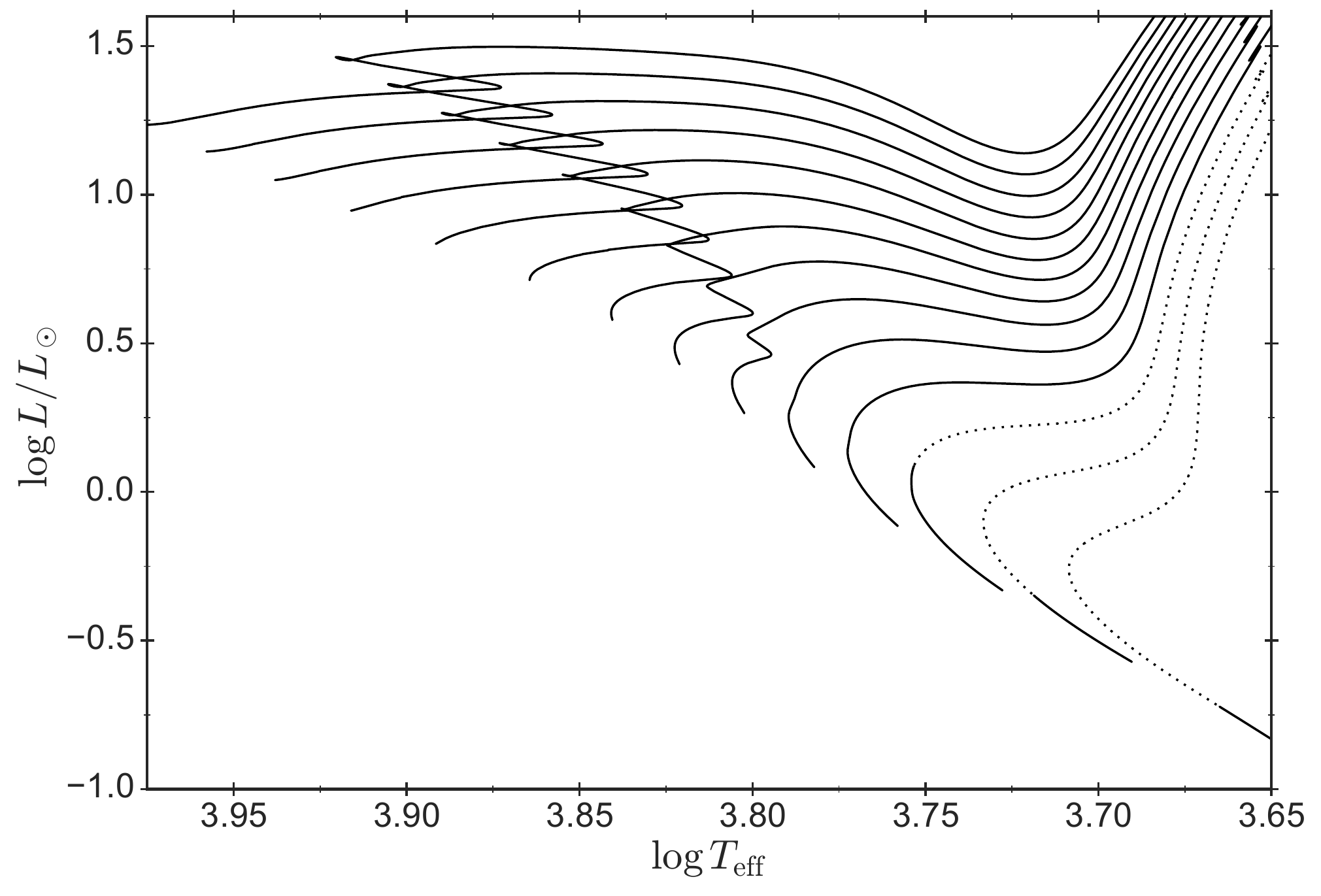}
\caption{Hertzsprung-Russell (or HR) diagram showing evolutionary tracks for masses between 0.7 and $2~\msun$ in steps of $0.1~\msun$. The portion of the tracks corresponding to ages greater than 14\,Gyr is shown as a dotted line.}
\label{fig:evol}
\end{figure}

%
%
Stellar evolution can be followed through numerical modelling, based on equations describing hydrostatic equilibrium, energy production and energy transport in stellar interiors, and resulting in a description of the changes in the surface and internal properties of the star as it ages. A comprehensive description of stellar evolution is provided by \citet{kippenhahn_etal_12}. The depletion of hydrogen in the core of main-sequence stars causes the core to contract, increasing the temperature in the stellar interior and consequently the surface luminosity, as well as the surface radius. When hydrogen is used up at the centre of the star nuclear fusion continues in a shell around the resulting helium core. In this phase the surface radius increases strongly, resulting in a decrease in the surface temperature; the star evolves towards the red-giant branch where the evolution continues at nearly constant surface temperature, as the star evolves towards strongly increasing luminosity with increasing radius (see Eq. \ref{eq:teff}). The evolution of stars is typically illustrated in a Hertzsprung-Russell (or HR) diagram, as shown in Fig. \ref{fig:evol}; an alternative, particularly when comparing with observations where the luminosity $L$ is unknown, is the so-called Kiel diagram, where luminosity is replaced by $\log\,g$.

\begin{figure}
\centering
\includegraphics[width=\textwidth]{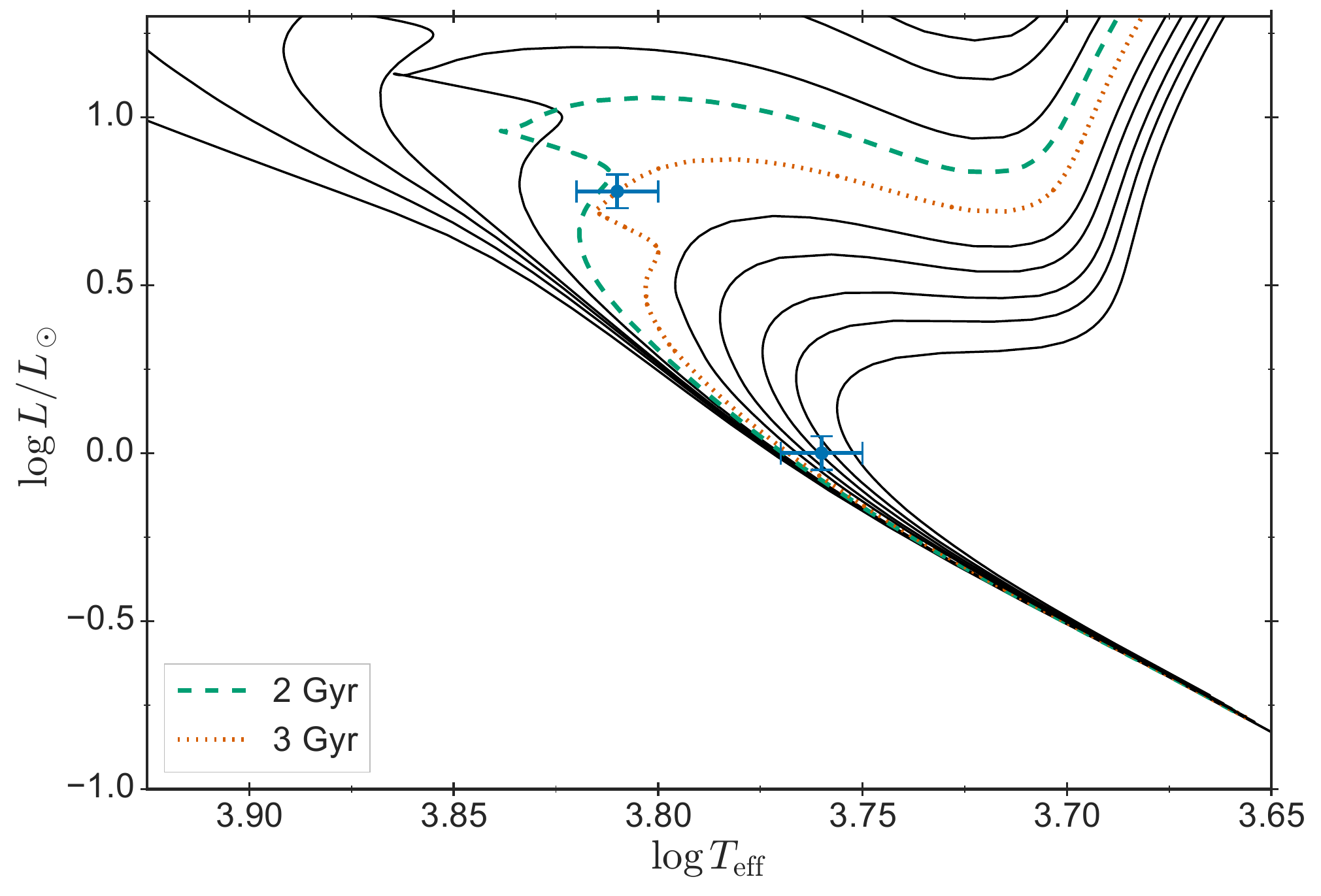}
\caption{Stellar isochrones at solar composition of ages (from left to right) 0, 0.5, 0.8, 1, 1.5, 2, 3, 4.6, 6, 8, 10, and 13\,Gyr. The symbols represent the position of the Sun and that of a $1.48\,\msun$ star at 2\,Gyr of age. Plotted errors correspond to 0.01 in $\log T_{\rm eff}$ and 0.1 in $\log L$. Dashed blue and dotted red lines depict the 2 and 3\,Gyr isochrones, respectively.}
\label{fig:iso}
\end{figure}

The relation between stellar observable properties and age is clearer in a plot of the so-called {\it isochrones} (see Fig. \ref{fig:iso}), which shows the location of models of a given age, but varying mass. The lowest curve marks the zero-age main sequence (ZAMS), where stars have just started hydrogen fusion in the core. With increasing age, stars of lower and lower mass have reached the end of the central hydrogen fusion, leading to the isochrones bending away from the ZAMS. The later parts of the isochrones, corresponding to phases where evolution is relatively rapid, closely resemble the evolution tracks in Fig. \ref{fig:evol}.

From Fig. \ref{fig:iso} it is in principle possible to determine the age of a star, given observational determination of its effective temperature and luminosity. The practical problems are illustrated by the two simulated observations marked in the figure, for a star roughly corresponding to the Sun and for a 2\,Gyr star of mass $\sim1.5\,\msun$, assuming realistic uncertainties in $T_{\rm eff}$ and $L$. For the lower-mass case it is evident that models of essentially any age are consistent with the observed $(T_{\rm eff}, L)$. In the higher-mass case a clearer age determination would be possible, although here there is some non-uniqueness concerning the evolutionary stage of the star. To these uncertainties must be added the errors in the observed surface composition of the star, which affects the location of the isochrones. It is evident that age determination, based on isochrones, is uncertain, particularly for stars at or below solar mass.

To sharpen the analysis, prior information may be included in age determination. An important aspect is the relative speed of evolution in the different evolutionary stages. In the example of the higher-mass star in Fig.~\ref{fig:iso} the choice of the 3\,Gyr isochrone would correspond to identifying the star to be in a phase of rapid evolution, where it would be less likely to be found. Also, one may take into account the fact that the fraction of stars increases rapidly with decreasing mass, reflecting the initial mass function. These aspects can be included through Bayesian analysis, where the assumed prior probability density is included in the estimate of final probability density function of, for example, the age \citep{pont_eyer_04, jorgensen_lindegren_05, serenelli_etal_13}.

\begin{figure}
\centering
\includegraphics[width=\textwidth]{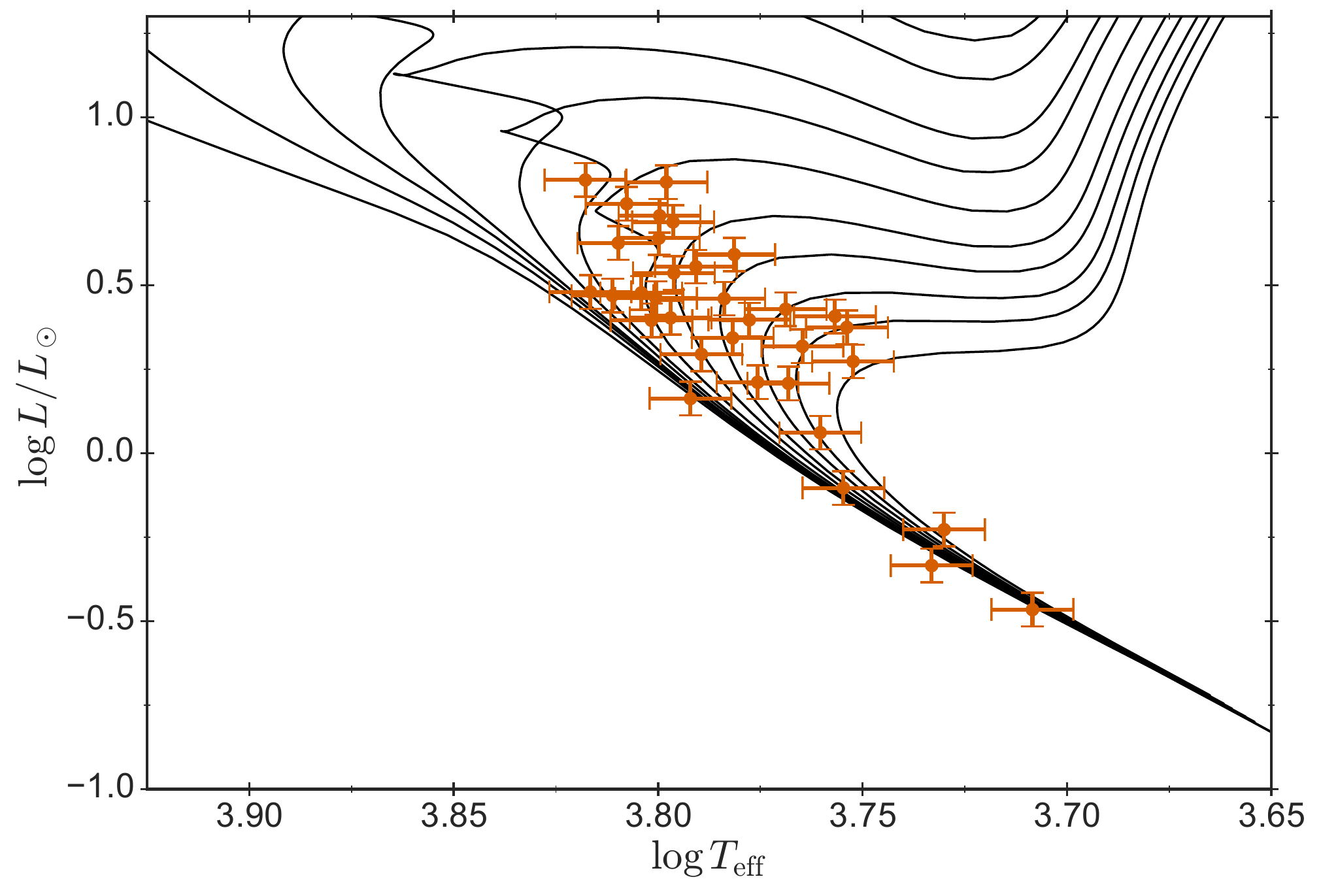}
\caption{Position in the HR diagram of the sample of 33 exo\-planet host stars analysed by \citet{silva_aguirre_etal_15}. Stellar isochrones are the same as those in Fig.~\ref{fig:iso}. Uncertainties in $\log T_{\rm eff}$ and $\log L$ correspond to 0.01 and 0.05, respectively. See text for details.}
\label{fig:kages_iso}
\end{figure}

An example portraying the difficulties in determining precise ages based on observed effective temperature and luminosity is shown in Fig.~\ref{fig:kages_iso}, depicting the position in the HR diagram of the sample of 33 {\it Kepler} exo\-planet hosts analysed by \citet{silva_aguirre_etal_15}. These targets were selected as they comprise the set of main-sequence stars harbouring planets with asteroseismic data of the highest quality that enables extraction of individual oscillation frequencies. As discussed in detail in the chapter {\it Characterizing host stars using asteroseismology} the inclusion of observed frequencies of solar-like oscillations in an asteroseismic analysis greatly constrains the inferred ages. Therefore, this set of stars serves as an excellent benchmark to compare the precision in age determination attainable from different methods.

\begin{figure}
\centering
\includegraphics[width=\textwidth]{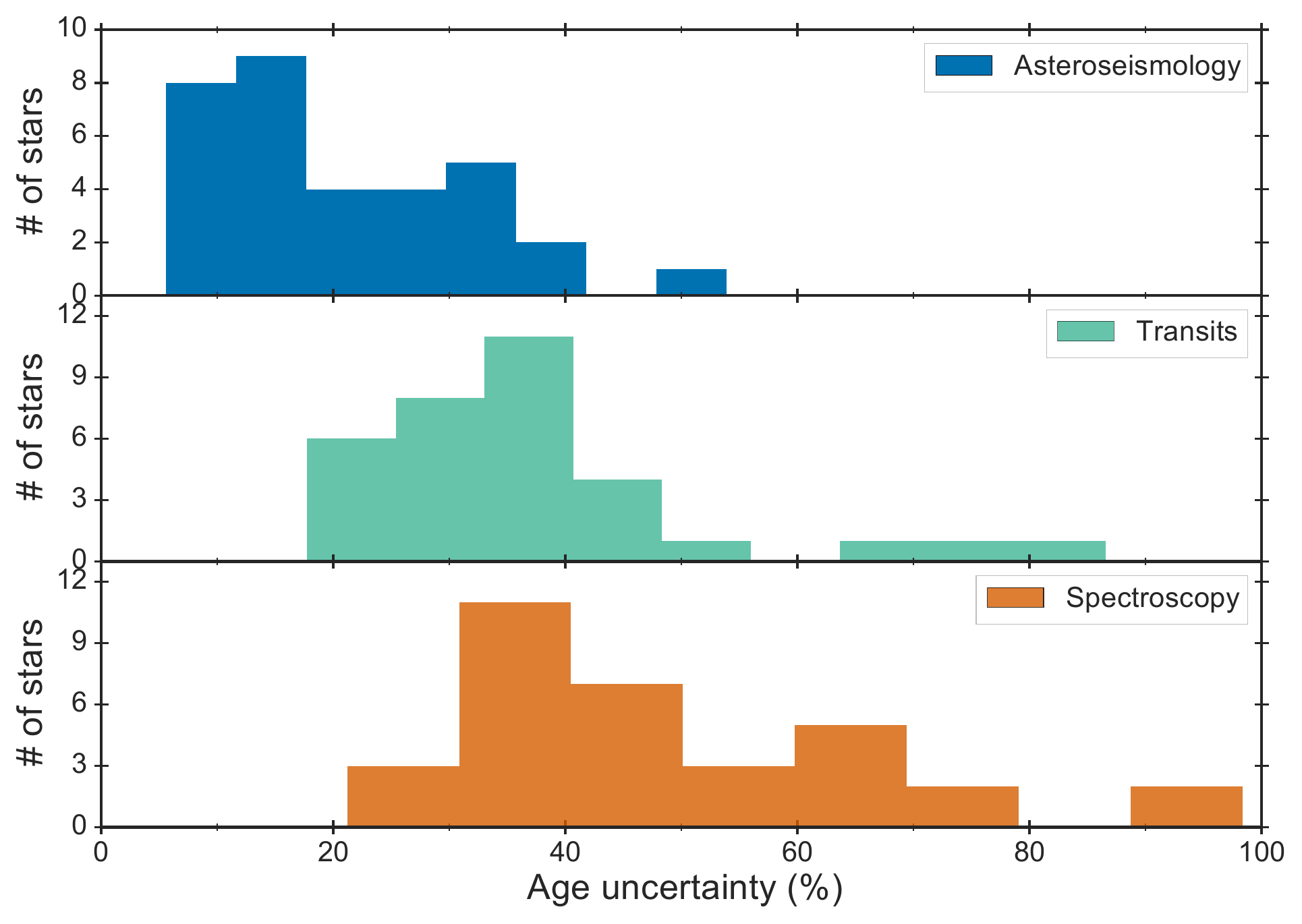}
\caption{Uncertainties in ages of exo\-planet host stars determined with {\tt BASTA} when fitting different sets of observational constraints. {\it Top panel:} individual oscillation frequencies, effective temperature, and metallicity. {\it Middle panel:} stellar density from the transit light curve, surface gravity, effective temperature, and metallicity.  {\it Bottom panel:} surface gravity, effective temperature, and metallicity. See text for details.}
\label{fig:kages}
\end{figure}

The three panels of Fig.~\ref{fig:kages} display the resulting age uncertainties in the sample of 33 stars when different sets of observational data are fit to a grid of stellar models using a state-of-the-art Bayesian analysis scheme \citep[{\tt BASTA},][]{silva_aguirre_etal_15,silva_aguirre_etal_17}. In the absence of asteroseismic information, one can determine stellar properties using photospheric observables such as effective temperature, composition, and surface gravity (labelled spectroscopy). For the special case of stars with transiting exo\-planets the analysis of the light curve can provide information about the mean density of the star, which adds very valuable constraints to the isochrone fit \citep{maxted_etal_15a}. If a strong constraint on the orbital eccentricity is available the stellar density can be extracted to a level of 5\% in the very best cases \citep[e.g.,][]{sandford_17}, slightly improving the age determination as shown in the middle panel of Fig.~\ref{fig:kages} (labelled transits).

By far the most precise stellar ages for field stars are those resulting from the inclusion of asteroseismic information, where in addition to the effective temperature and composition the individual oscillation frequencies are fitted in {\tt BASTA} (labelled asteroseismology in Fig.~\ref{fig:kages}) . It is evident that for a large fraction of stars the age can be determined with a precision better than 20 \%. Specific examples of interesting precise age determinations are the age of $10.4 \pm 1.4$\,Gyr found for Kepler 10 \citep{fogtmann_etal_14} and the age of $11.2 \pm 1.0$\,Gyr for the five-planet system around Kepler 444 \citep{campante_etal_15}.These results show that planetary systems were formed early in the history of the Milky Way Galaxy.

A special case of isochrone fits concerns stellar clusters. In such clusters the stars are normally assumed to have been formed at the same time and with the same initial composition, while covering a range of masses; hence they all have the same age, such that the distribution of stars in the  $(T_{\rm eff}, L)$ diagram should correspond to that of an isochrone. Fitting an isochrone, in particular the turn-off from the ZAMS, to the observations therefore provides a determination of the age of the cluster without the uncertainties associated with the fits of individual stars. Thanks to the K2 extension of the {\it Kepler} mission exo\-planets  have been found in the Hyades and Praesepe stellar clusters, both with an age of around 800\,Myr\footnote{According to \citet{brandt_huang_15}; more conventional ages are around 600\,Myr for both clusters.} {\rd \citep{mann_etal_16a, mann_etal_17a, mann_etal_18}}, and in the very young Upper Scorpius OB Association \citep{mann_etal_16b}. This provides important information about the evolution of planetary systems with age. Also, well-determined ages of clusters play an important role in the calibration of other age-determination techniques. We return to this below.

These methods for age determination all rely on the modelling of stellar evolution and hence are sensitive to errors in the modelling. The basic theory, including the relevant physics of stellar matter, is probably fairly well established. However, there are serious uncertainties related to hydrodynamical instabilities and the resulting potential mixing processes. By changing the evolution of the core composition these would directly affect the relation between the age and the observable properties of the star. Main-sequence stars more massive than the Sun have a fully mixed convective core; a very uncertain aspect is extent of motion, the so-called convective overshoot, beyond the limit of convective instability, which may have important effects on stellar evolution. Constraints on overshooting have been obtained from asteroseismic analyses of main-sequence stars \citep{silva_aguirre_etal_13,deheuvels_etal_16} as well as eclipsing binaries \citep{claret16}, offering hope for a better understanding of this contribution to the age uncertainty. Mixing may also be induced by instabilities associated with the evolution of stellar rotation (see below).
\section{Ages from rotation and activity}
The rate of stellar rotation changes as stars evolve, making measurement of stellar rotation a potential age indicator. Stars are born as rapid rotators. In relatively cool stars with outer convection zones, supporting magnetic activity {\rd which is assumed to arise through some dynamo action and hence to be strongly linked to rotation}, angular momentum is lost to a magnetic stellar wind which is coupled to the convection zone. This leads to a reduction of the surface rotation rate. Coupling to the radiative interior through processes that are not fully understood reduces the angular momentum of the interior, leading at least in the solar case to roughly uniform rotation \citep[e.g.,][]{thompson_etal_03}.
{\rd As reviewed by \citet{jeffries_14} the evolution of rotation and hence magnetic activity with age provides potential for age determination.}

\begin{figure}
\centering
\includegraphics[width=\textwidth]{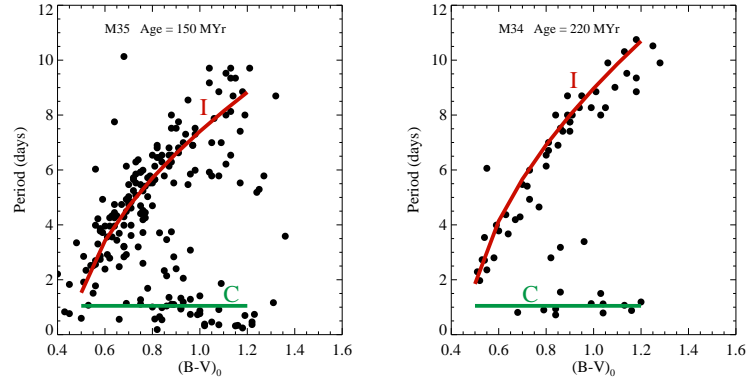}
\caption{Rotation periods as a function of (B-V) colour for stars belonging to two open clusters. The $C$ and $I$ sequences correspond to {\it Convective} and {\it Interface} magnetic fields as defined by \citet{barnes_03}. From \citet{brown_14}, using data from \citet{meibom_etal_09, meibom_etal_11b} \copyright AAS. Reproduced with permission.}
\label{fig:gyro}
\end{figure}

Based on a small number of rotation determinations \citet{skumanich_72} inferred that the rotation rate decreases with the age $t$ of the star as $t^{-1/2}$, with a corresponding decrease in stellar activity. \citet{durney_72} independently carried out a simplified analysis of the angular-momentum loss, leading to the same time dependence. Thus a determination of stellar rotation rates, with proper calibration, can be used for age determination. This is the basis for {\it gyrochronology}\footnote{ Note that since, as shown in Fig. \ref{fig:gyro}, the rotation periods also depend on the colour of the stars, a more proper term would be {\it gyrochromochronology}.} \citep{barnes_03, barnes_07}.

As discussed above, stellar clusters provide samples of stars with the same (and in many cases well-estimated) age and hence are an excellent basis for calibrating techniques for age determination. Figure \ref{fig:gyro} shows rotation periods for two clusters of different age plotted against colour index; this is a measure of the effective temperature and hence, at the cluster age, a measure of stellar mass, with the colour index increasing with decreasing mass. It is evident that the bulk of the stars in each case lie near a well-defined curve, with rotation period increasing with increasing $B - V$ and hence decreasing mass. The interpretation is that the loss of angular momentum through the magnetic stellar wind leads to a well-defined relation between age and rotation period, but at a rate that depends on the mass. The relatively few rapidly-rotating stars may yet have to reach this relation, possibly because of very rapid initial rotation or a different configuration of their magnetic field \citep{brown_14}.

Two main observational techniques are available for the determination of stellar rotation periods. Rotation contributes to the line broadening through the Doppler effect due to the line-of-sight component of the rotational velocity $v_{\rm rot}$; thus only $v_{\rm rot} \sin i$ is determined, where $i$ is the inclination of the rotation axis relative to the line of sight. Stellar rotation can also in many cases be determined from periodic variations in the luminosity of spotted stars. Nevertheless, the extraction of rotation periods from both the width of spectral lines and from the light modulation induced by starspots becomes increasingly difficult with increasing age and hence increasing rotation period. The rotational broadening of the spectral lines becomes comparable with the natural line width and hence difficult to isolate, and the decreasing magnetic activity level decreases the magnitude of the starspot modulation of the stellar intensity.

\begin{figure}
\centering
\includegraphics[width=\textwidth]{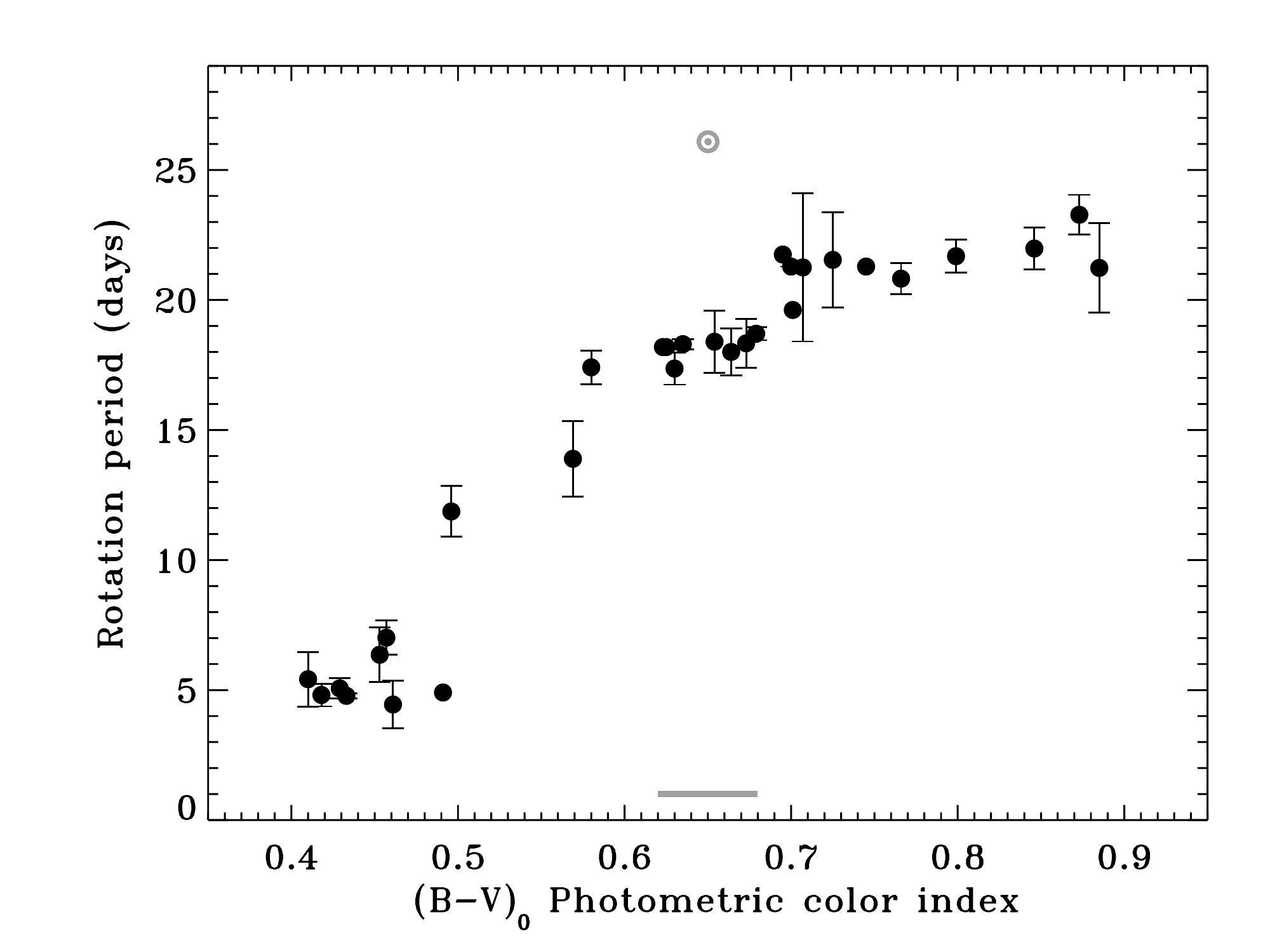}
\caption{Rotation periods as a function of (B-V) photometric index for a sample of stars in the open cluster  NGC\,6819 observed by {\it Kepler}. Stellar masses are given in the top of the figure while the position of the Sun in this diagram is shown by the solar symbol. From \citet{meibom_etal_15}, \copyright Nature. Reproduced with permission.}
\label{fig:6819}
\end{figure}

\begin{figure}
\centering
\includegraphics[width=\textwidth]{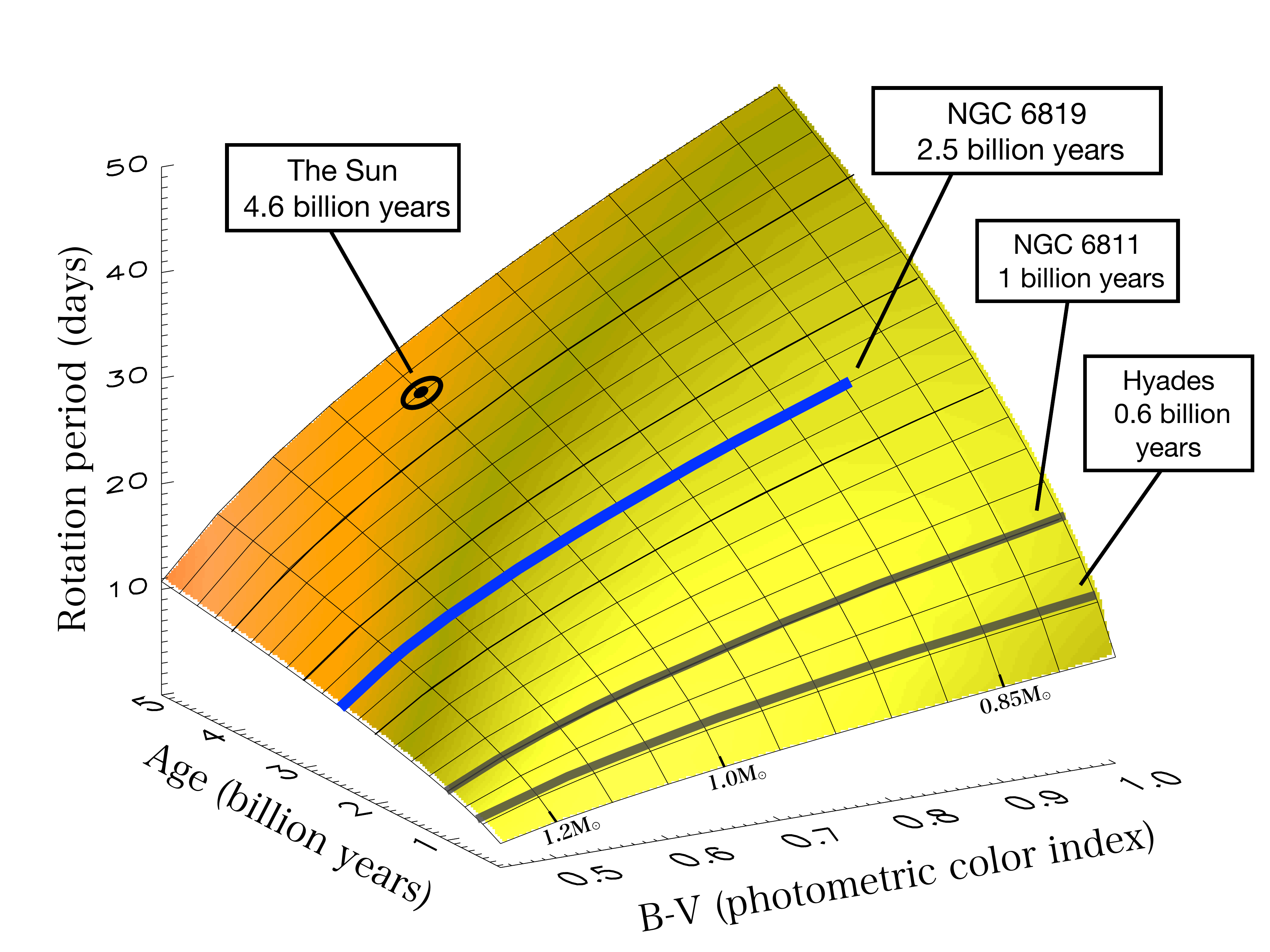}
\caption{Predicted surface depicting the relation between period, age, and colour index extrapolated to ages greater than that of NGC\,6819 by anchoring it to the Sun. From \citet{meibom_etal_15}, \copyright Nature. Reproduced with permission.}
\label{fig:meirot}
\end{figure}

The high precision of space photometry, in particular with {\it Kepler}, has allowed the determination of rotation periods for a wide variety of stars. In particular, \citet{meibom_etal_11, meibom_etal_15} extracted rotation periods for stars in the two clusters NGC\,6811 and NGC\,6819, of age 2.1 and 2.5\,Gyr, respectively. Figure~\ref{fig:6819} illustrates the close relation found between colour and rotation period for NGC\,6819. On this basis \citet{meibom_etal_15} could extend the $t^{-1/2}$ dependence of rotation on age \citep{skumanich_72} to a broader range, as illustrated schematically in Fig~\ref{fig:meirot}, where the Sun is used to anchor the relation at 4.57\,Gyr. The relation was further extended by \citet{barnes_etal_16} to nearly solar age using K2 observations of the 4\,Gyr old cluster M\,67.

\begin{figure}
\centering
\includegraphics[width=\textwidth]{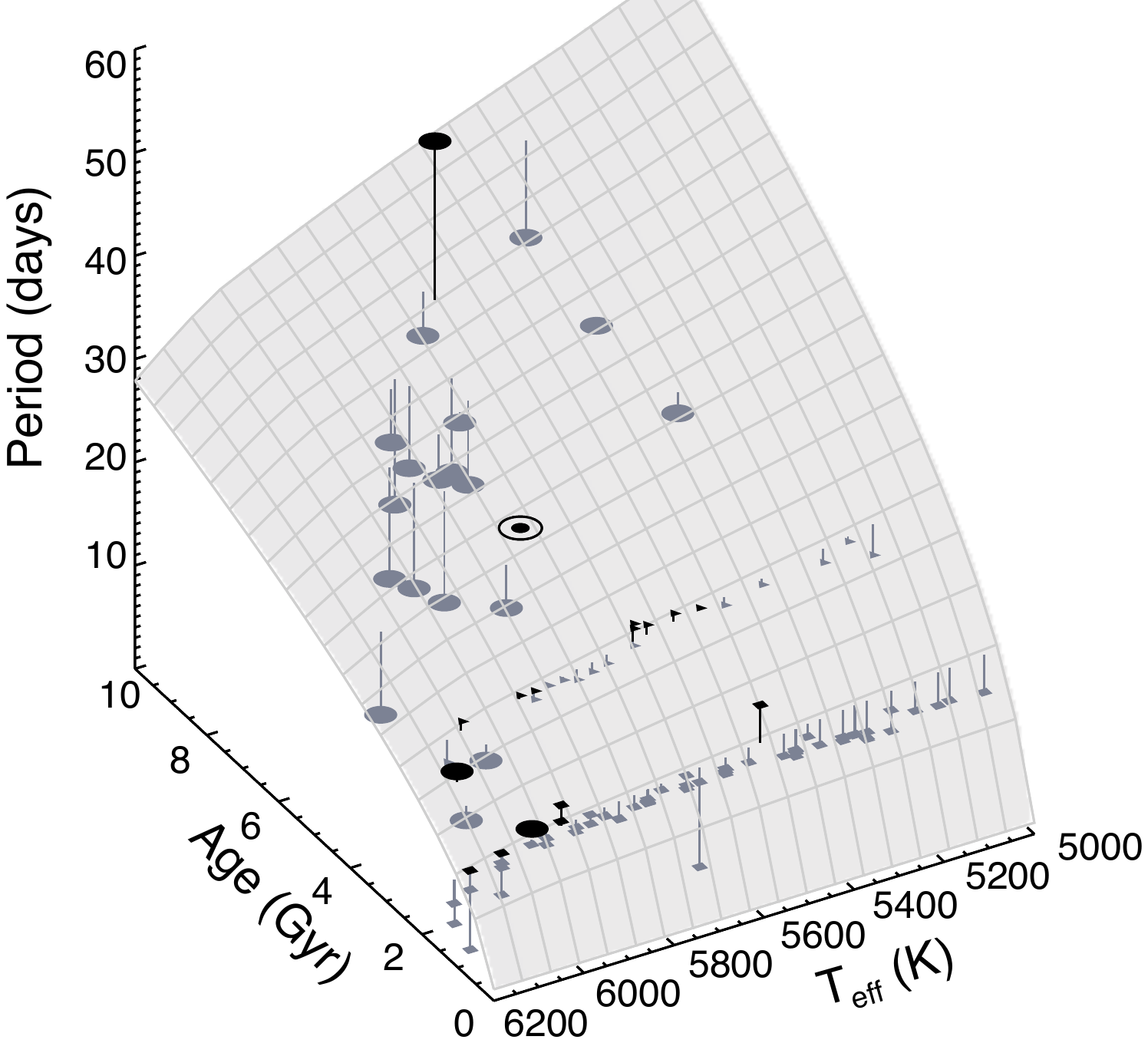}
\vskip -7cm
\caption{Similar to Fig.~\ref{fig:meirot} where asteroseismic ages of field stars with rotation periods measured from {\it Kepler} data are represented by filled circles. Squares and triangles depict the open clusters NGC\,6811 and NGC\,6819, respectively. From \citet{van_saders_etal_16}, \copyright Nature. Reproduced with permission.}
\label{fig:vansad}
\end{figure}

Few open clusters are older than M\,67, 
and for these determinations of rotation periods have not been possible. However, the {\it Kepler} observations have provided rotation-period determinations and asteroseismic age inferences for a number of stars, many of which are older \citep{van_saders_etal_16}. The results are illustrated in Fig. \ref{fig:vansad}, comparing them with the fit provided by \citet{meibom_etal_15}. Strikingly, there are very substantial deviations, with stars older than the Sun in most cases rotating considerably faster than predicted by the simple $t^{-1/2}$ variation, indicating a decrease in the efficiency of the angular-momentum loss. \citet{van_saders_etal_16} speculate that this is related to the evolution with age of the so-called Rossby number Ro, i.e., the ratio between the rotation period and the convective turn-over timescale, and perhaps related to a change of the magnetic topology from being predominantly dipolar to a more complex structure. It appears that the departure from the simple age dependence sets in when Ro exceeds a critical value of around 2; interestingly, the Sun is close to this limit. These issues are extremely interesting in the context of the origin and evolution of stellar magnetic activity, but for age determination based on gyrochronology they are clearly a complication which restricts the reliable use to relatively young stars, at least until we have a better understanding of the transition to weaker angular-momentum loss and its relation to observable stellar properties.

Given the close relation between stellar rotation and magnetic activity, observations sensitive to activity can also be used for age diagnostics. A commonly used index is obtained from the chromospheric emission in the cores of the singly-ionized calcium H and K lines, which are located in the near-ultraviolet. As shown by \citet{skumanich_72} this emission decreases with age, like the rotation, as $t^{-1/2}$. \citet{mamajek_hillenbrand_07} made a careful analysis of the relation between age and the calcium index, considering also combining this with the information provided by the rotation rate. They estimated that the achievable precision when using these techniques in $\log t$ is around 0.1 -- 0.2, in the age range 0.6 -- 4.5\,Gyr, with worse precision for younger and older stars. Indeed, one might have expected that the departures from the Skumanich relation for rotation found by \citet{van_saders_etal_16} in older stars would also affect the relation between age and magnetic activity. Stellar activity, inferred from X-ray observations of stellar coronae, was compared with asteroseismic ages by \citet{booth_etal_17}. Interestingly, they found that the activity decreased more rapidly with age for stars older than the Sun than for younger stars, unlike the slower decrease of rotation with age inferred by  \citet{van_saders_etal_16}. This suggests a more complex relation between rotation and activity than previously assumed, clearly further complicating the use of rotation and activity as age indicators.
%
%
%
\section{Ages from lithium abundance} 
%
%
Lithium is created in big-bang nucleosynthesis and is destroyed by nuclear reactions at relatively modest temperature; in the case of the more abundant ${}^7 {\rm Li}$ isotope, which we consider in the following, destruction on a comparatively short timescale happens for temperatures above $2.5 \times 10^6$\,K. In the solar case the surface lithium abundance has been reduced by roughly a factor 150 compared with the assumed initial abundance. Thus the evolution with time of the surface lithium abundance provides a potential age diagnostics
{\rd \citep[see][for a review]{jeffries_14}}. \citet{sestito_randich_05}, for example, analysed lithium observations for a number of stellar clusters to relate the abundance to age. However, the details of the processes that lead to lithium destruction are uncertain. In general, the outer convective envelope is not sufficiently deep to reach the temperatures required for lithium destruction. Thus, additional mixing below the convectively unstable region is required. This is likely related to the angular momentum transport that couples the decrease in the surface rotation rate, owing to the angular-momentum loss to a magnetic stellar wind, to the deeper interior of the star \citep[e.g.,][]{somers_pinsonneault_16}. It appears that most of the lithium destruction takes place during the first few Gyr of the evolution, restricting the usefulness of lithium as an age diagnostics to relatively young stars.
\section{Conclusions}
%
%
In this chapter we have discussed various non-asteroseismic techniques for age determination in {\rd lower-main-sequence} stars which are the most typical exo\-planet hosts. None of these techniques can compete with asteroseismology, based on individually determined frequencies, in terms of precision and likely accuracy. On the other hand, with present and foreseeable observational data detailed asteroseismology will only be possible for a fraction of the stars for which exo\-planets are found. Thus alternative techniques are essential. Stellar evolution and hence age is directly reflected by the location of the star in the HR diagram, allowing age determination through isochrone fitting, although with strong degeneracy, particularly for low-mass stars. Including the mean stellar density, as may be determined from transiting 
exo\-planets, improves the resolution in the best cases when a strong density prior is available. Gyrochronology, based on the relation between rotation period, colour and age, is promising, at least for stars up to solar age. For exo\-planets detected with the transit technique the determination of the rotation period from spot modulation will in many cases be a natural by-product of the observations. For older stars, with low activity, this signal becomes weaker and possibly hard to interpret, and in addition the relation between age and rotation period appears to undergo a transition which is so far not fully understood.

To test the validity of age inferences, comparisons of the results of different techniques are clearly very valuable. In the special case of isochrone fitting and gyrochronology for exo\-planet hosts there seems to be some inconsistency, with gyrochronology in some cases indicating substantially lower ages, corresponding to faster rotation \citep[e.g.,][and references therein]{maxted_etal_15b}. One proposed origin for this discrepancy, at least for stars with hot Jupiters, is that the rotation of the star has been spun up through tidal interaction with the close-in planet, which would clearly affect gyrochronology. Further studies are required to investigate this;  it is obvious that asteroseismic ages for such systems would be of great value, to resolve the apparent conflict.

Very cool and low-mass main-sequence stars, the so-called M dwarfs, are of particular interest for exo\-planet research: their small radii and masses make earth-size planets easier to detect, and their very low luminosities place the habitable zone close to the star, with orbital periods of only a couple of weeks, further increasing the chances of finding relevant planets. A remarkable example is the Trappist-1 system, with three potentially habitable planets \citep{gillon_etal_17}. Age determination for these systems is difficult. The nuclear evolution of such low-mass stars is extremely slow, such that the location of the star in the HR diagram provides little constraint on the age. Also, the predicted amplitudes of solar-like oscillations are very low, precluding asteroseismology. \citet{burgasser_etal_17} carried out an extensive analysis of Trappist-1, including location in the HR diagram, mean density, rotation and lithium abundance, to infer an age of $7.6 \pm 2.2$\,Gyr. Further work on ages of very low mass stars, e.g., based on stars in clusters with well-determined ages, would be very valuable.

Age determination based on nuclear evolution, including asteroseismology, depends directly on the reduction of the hydrogen abundance in the stellar core and hence is  insensitive to the very interesting early phases of the life of a star, where nuclear burning is just starting. Here the variations in stellar rotation and activity may provide age  information, although with very substantial uncertainties owing to a strong dependence on initial conditions and our incomplete understanding of the relevant processes. Further constraints may be available from lithium destruction, although again limited by uncertainty about the relevant mixing processes. For pre-main-sequence stars in the $\delta$ Scuti instability strip \citet{zwintz_etal_14} demonstrated a clear relation between the age and the oscillation properties, largely reflecting the changing radii of the stars as they contract towards the main sequence. This has the potential for further development towards a more sensitive technique for age determination in this case.
%

We emphasize again that the accuracy of any technique for age determination is limited by our incomplete understanding of the underlying processes, and, obviously, by the errors in the observations underlying the determination. It is fair to say that the nuclear evolution, and hence age determination depending on it, is quite well understood in broad outline, although substantial uncertainties remain, particularly regarding mixing processes in the stellar interior which directly affect the rate at which hydrogen is consumed. Here improvements will follow from more detailed asteroseismic investigations of stellar interiors, together with an improved understanding of the relevant hydrodynamical processes leading to mixing and implementation of this understanding in stellar-evolution calculations. It should be kept in mind that the solar case offers by far the best  potential for detailed comparisons between models and observations; the fact that the most recent solar models, using updated surface abundances, fail this test is a serious concern with potential implications for other stars. This is also related to the effects of other uncertainties in the required determination of stellar composition, although these concerns are probably most relevant for the intrinsically precise asteroseismic determinations.

For other age determination techniques considered here the limitations in our knowledge about the underlying physical processes are substantially bigger. This is true for gyrochronology, where the rate of angular momentum loss from the outer layers and the coupling to the rotation in the interior are poorly understood. Similar uncertainties affect the use of stellar activity as an age indicator, given that even the magnetic activity of the Sun and its relation to the properties of rotation and convection are far from fully understood. More detailed modelling of these processes is obviously important, and further insight will be obtained from detailed observations of the surface magnetic configuration for stars of different age {\rd \citep[e.g.][]{folsom_etal_16, folsom_etal_18}}. However, improvement of the reliability of rotation and activity as age indicators will also depend strongly on further calibration against other age determinations, in particular open-cluster fitting and asteroseismology. The understanding of the evolution of rotation also affects the use of lithium destruction as an age indicator, given the very strong sensitivity of lithium burning to the mixing processes in the stellar interior.

Needless to say, these developments require analysis of additional data. Much data remain to be studied in detail from the nominal {\it Kepler} mission, which for a considerable time will remain amongst the best available data. Also, new missions are on the way combining exo\-planet research  and stellar astrophysics. The TESS mission \citep{ricker_etal_16}, to be launched in the spring of 2018, will provide data for a huge number of stars, although for shorter periods. Since these are typically brighter than the {\it Kepler} targets, there is a potential for more extensive follow-up observations, e.g., concerning the effects of rotation and chemical composition. Also, the PLATO mission \citep{rauer_etal_14}, with scheduled launch in 2026, will provide observations of a quality matching {\it Kepler} but for a much larger number of brighter stars, combining long time series and excellent opportunities for follow-up observations. This promises a breakthrough in the characterization of exo\-planet hosts, including the determination of ages. To make the full use of this potential the analysis of the PLATO data must be organized to combine the exo\-planet characterization with improvements in stellar modelling, concerning all the aspects discussed here, on the basis of the PLATO observations.
\section{Cross-References}
\begin{itemize}
\item{Characterizing host stars using asteroseismology}
\item{Rotation of Planet-Hosting Stars}
\end{itemize}

\begin{acknowledgement}
Funding for the Stellar Astrophysics Centre is provided by The Danish National Research Foundation (Grant DNRF106). V.S.A. acknowledges support from the Villum Foundation (Research grant 10118).
\end{acknowledgement}


\begin{thebibliography}{57}
\providecommand{\natexlab}[1]{#1}
\providecommand{\url}[1]{{#1}}
\providecommand{\urlprefix}{URL }
\expandafter\ifx\csname urlstyle\endcsname\relax
  \providecommand{\doi}[1]{DOI~\discretionary{}{}{}#1}\else
  \providecommand{\doi}{DOI~\discretionary{}{}{}\begingroup
  \urlstyle{rm}\Url}\fi
\providecommand{\eprint}[2][]{\url{#2}}

\bibitem[{{Asplund} et~al.(2009){Asplund}, {Grevesse}, {Sauval}, and
  {Scott}}]{asplund_etal_09}
{Asplund} M, {Grevesse} N, {Sauval} AJ {Scott} P (2009) {The Chemical
  Composition of the Sun}. \araa 47:481--522

\bibitem[{{Barnes}(2003)}]{barnes_03}
{Barnes} SA (2003) {On the Rotational Evolution of Solar- and Late-Type Stars,
  Its Magnetic Origins, and the Possibility of Stellar Gyrochronology}. \apj
  586:464--479

\bibitem[{{Barnes}(2007)}]{barnes_07}
{Barnes} SA (2007) {Ages for Illustrative Field Stars Using Gyrochronology:
  Viability, Limitations, and Errors}. \apj 669:1167--1189

\bibitem[{{Barnes} et~al.(2016){Barnes}, {Weingrill}, {Fritzewski},
  {Strassmeier}, and {Platais}}]{barnes_etal_16}
{Barnes} SA, {Weingrill} J, {Fritzewski} D, {Strassmeier} KG {Platais} I (2016)
  {Rotation Periods for Cool Stars in the 4 Gyr old Open Cluster M67, The
  Solar-Stellar Connection, and the Applicability of Gyrochronology to at least
  Solar Age}. \apj 823:16

\bibitem[{{Basu} and {Antia}(2008)}]{basu_antia_08}
{Basu} S {Antia} HM (2008) {Helioseismology and solar abundances}. {\rm Phys
  Rep} 457:217--283

\bibitem[{{Booth} et~al.(2017){Booth}, {Poppenhaeger}, {Watson}, {Silva
  Aguirre}, and {Wolk}}]{booth_etal_17}
{Booth} RS, {Poppenhaeger} K, {Watson} CA, {Silva Aguirre} V {Wolk} SJ (2017)
  {An improved age-activity relationship for cool stars older than a gigayear}.
  \mnras 471:1012--1025

\bibitem[{Boyajian et~al.(2012)Boyajian, McAlister, van Belle, Gies, ten
  Brummelaar, von Braun, Farrington, Goldfinger, O'Brien, Parks, Richardson,
  Ridgway, Schaefer, Sturmann, Sturmann, Touhami, Turner, and
  White}]{Boyajian_2012}
Boyajian TS, McAlister HA, van Belle G et~al. (2012) {Stellar diameters and
  temperatures. I. Main-sequence A, F, and G stars}. ApJ 746(1):101

\bibitem[{{Brandt} and {Huang}(2015)}]{brandt_huang_15}
{Brandt} TD {Huang} CX (2015) {The Age and Age Spread of the Praesepe and
  Hyades Clusters: a Consistent, \~{}800 Myr Picture from Rotating Stellar
  Models}. \apj 807:24

\bibitem[{{Brown}(2014)}]{brown_14}
{Brown} TM (2014) {The Metastable Dynamo Model of Stellar Rotational
  Evolution}. \apj 789:101

\bibitem[{{Burgasser} and {Mamajek}(2017)}]{burgasser_etal_17}
{Burgasser} AJ {Mamajek} EE (2017) {On the Age of the TRAPPIST-1 System}. \apj
  845:110

\bibitem[{{Campante} et~al.(2015){Campante}, {Barclay}, {Swift}, {Huber},
  {Adibekyan}, {Cochran}, {Burke}, {Isaacson}, {Quintana}, {Davies}, {Silva
  Aguirre}, {Ragozzine}, {Riddle}, {Baranec}, {Basu}, {Chaplin},
  {Christensen-Dalsgaard}, {Metcalfe}, {Bedding}, {Handberg}, {Stello},
  {Brewer}, {Hekker}, {Karoff}, {Kolbl}, {Law}, {Lundkvist}, {Miglio}, {Rowe},
  {Santos}, {Van Laerhoven}, {Arentoft}, {Elsworth}, {Fischer}, {Kawaler},
  {Kjeldsen}, {Lund}, {Marcy}, {Sousa}, {Sozzetti}, and
  {White}}]{campante_etal_15}
{Campante} TL, {Barclay} T, {Swift} JJ et~al. (2015) {An Ancient Extrasolar
  System with Five Sub-Earth-size Planets}. \apj 799:170

\bibitem[{Casagrande et~al.(2014)Casagrande, Portinari, Glass, Laney,
  Silva~Aguirre, Datson, Andersen, Nordstr{\"o}m, Holmberg, Flynn, and
  Asplund}]{Casagrande_2014}
Casagrande L, Portinari L, Glass IS et~al. (2014) {Towards stellar effective
  temperatures and diameters at 1 per cent accuracy for future surveys}.
  Monthly Notices of the Royal Astronomical Society 439(2):2060--2073

\bibitem[{{Chromey}(2016)}]{chromey_16}
{Chromey} FR (2016) {To Measure the Sky}

\bibitem[{{Claret} and {Torres}(2016)}]{claret16}
{Claret} A {Torres} G (2016) {The dependence of convective core overshooting on
  stellar mass}. \aap 592:A15

\bibitem[{{Connelly} et~al.(2012){Connelly}, {Bizzarro}, {Krot}, {Nordlund},
  {Wielandt}, and {Ivanova}}]{connelly_etal_12}
{Connelly} JN, {Bizzarro} M, {Krot} AN et~al. (2012) {The Absolute Chronology
  and Thermal Processing of Solids in the Solar Protoplanetary Disk}. Science
  338:651

\bibitem[{{Deheuvels} et~al.(2016){Deheuvels}, {Brand{\~a}o}, {Silva Aguirre},
  {Ballot}, {Michel}, {Cunha}, {Lebreton}, and
  {Appourchaux}}]{deheuvels_etal_16}
{Deheuvels} S, {Brand{\~a}o} I, {Silva Aguirre} V et~al. (2016) {Measuring the
  extent of convective cores in low-mass stars using Kepler data: toward a
  calibration of core overshooting}. \aap 589:A93

\bibitem[{{Durney}(1972)}]{durney_72}
{Durney} B (1972) {Evidence for Changes in the Angular Velocity of the Surface
  Regions of the Sun and Stars - Comments}. NASA Special Publication 308:282

\bibitem[{{Fogtmann-Schulz} et~al.(2014){Fogtmann-Schulz}, {Hinrup}, {Van
  Eylen}, {Christensen-Dalsgaard}, {Kjeldsen}, {Silva Aguirre}, and
  {Tingley}}]{fogtmann_etal_14}
{Fogtmann-Schulz} A, {Hinrup} B, {Van Eylen} V et~al. (2014) {Accurate
  Parameters of the Oldest Known Rocky-exoplanet Hosting System: Kepler-10
  Revisited}. \apj 781:67

\bibitem[{{Folsom} et~al.(2016){Folsom}, {Petit}, {Bouvier}, {L{\`e}bre},
  {Amard}, {Palacios}, {Morin}, {Donati}, {Jeffers}, {Marsden}, and
  {Vidotto}}]{folsom_etal_16}
{Folsom} CP, {Petit} P, {Bouvier} J et~al. (2016) {The evolution of surface
  magnetic fields in young solar-type stars - I. The first 250 Myr}. \mnras
  457:580--607

\bibitem[{{Folsom} et~al.(2018){Folsom}, {Bouvier}, {Petit}, {L{\`e}bre},
  {Amard}, {Palacios}, {Morin}, {Donati}, and {Vidotto}}]{folsom_etal_18}
{Folsom} CP, {Bouvier} J, {Petit} P et~al. (2018) {The evolution of surface
  magnetic fields in young solar-type stars II: the early main sequence
  (250-650 Myr)}. \mnras 474:4956--4987

\bibitem[{{Gaia Collaboration} et~al.(2016){Gaia Collaboration}, Brown,
  Vallenari, Prusti, de~Bruijne, Mignard, Drimmel, Babusiaux, Bailer-Jones,
  Bastian, Biermann, Evans, Eyer, Jansen, Jordi, Katz, Klioner, Lammers,
  Lindegren, Luri, O'Mullane, Panem, Pourbaix, Randich, Sartoretti, Siddiqui,
  Soubiran, Valette, Van~Leeuwen, Walton, Aerts, Arenou, Cropper, H{\o}g,
  Lattanzi, Grebel, Holland, Huc, Passot, Perryman, Bramante, Cacciari,
  Casta{\~n}eda, Chaoul, Cheek, De~Angeli, Fabricius, Guerra, Hern{\'a}ndez,
  Jean-Antoine-Piccolo, Masana, Messineo, Mowlavi, Nienartowicz,
  Ord{\'o}{\~n}ez-Blanco, Panuzzo, Portell, Richards, Riello, Seabroke, Tanga,
  Th{\'e}venin, Torra, Els, Gracia-Abril, Comoretto, Garcia-Reinaldos, Lock,
  Mercier, Altmann, Andrae, Astraatmadja, Bellas-Velidis, Benson, Berthier,
  Blomme, Busso, Carry, Cellino, Clementini, Cowell, Creevey, Cuypers,
  Davidson, De~Ridder, de~Torres, Delchambre, Dell'Oro, Ducourant, Fr{\'e}mat,
  Garc{\'\i}a-Torres, Gosset, Halbwachs, Hambly, Harrison, Hauser, Hestroffer,
  Hodgkin, Huckle, Hutton, Jasniewicz, Jordan, Kontizas, Korn, Lanzafame,
  Manteiga, Moitinho, Muinonen, Osinde, Pancino, Pauwels, Petit, Recio-Blanco,
  Robin, Sarro, Siopis, Smith, Smith, Sozzetti, Thuillot, van Reeven, Viala,
  Abbas, Abreu~Aramburu, Accart, Aguado, Allan, Allasia, Altavilla,
  {\'A}lvarez, Alves, Anderson, Andrei, Anglada~Varela, Antiche, Antoja, Anton,
  Arcay, Bach, Baker, Balaguer-N{\'u}{\~n}ez, Barache, Barata, Barbier,
  Barblan, Barrado~y Navascu{\'e}s, Barros, Barstow, Becciani, Bellazzini,
  Bello~Garc{\'\i}a, Belokurov, Bendjoya, Berihuete, Bianchi, Bienaym{\'e},
  Billebaud, Blagorodnova, Blanco-Cuaresma, Boch, Bombrun, Borrachero,
  Bouquillon, Bourda, Bouy, Bragaglia, Breddels, Brouillet, Br{\"u}semeister,
  Bucciarelli, Burgess, Burgon, Burlacu, Busonero, Buzzi, Caffau, Cambras,
  Campbell, Cancelliere, Cantat-Gaudin, Carlucci, Carrasco, Castellani,
  Charlot, Charnas, Chiavassa, Clotet, Cocozza, Collins, Costigan, Crifo,
  Cross, Crosta, Crowley, Dafonte, Damerdji, Dapergolas, David, David, De~Cat,
  de~Felice, de~Laverny, De~Luise, De~March, de~Martino, de~Souza, Debosscher,
  del Pozo, Delbo, Delgado, Delgado, Di~Matteo, Diakite, Distefano, Dolding,
  Dos~Anjos, Drazinos, Duran, Dzigan, Edvardsson, Enke, Evans, Eynard~Bontemps,
  Fabre, Fabrizio, Faigler, Falc{\~a}o, Farr{\`a}s~Casas, Federici, Fedorets,
  Fern{\'a}ndez-Hern{\'a}ndez, Fernique, Fienga, Figueras, Filippi, Findeisen,
  Fonti, Fouesneau, Fraile, Fraser, Fuchs, Gai, Galleti, Galluccio, Garabato,
  Garc{\'\i}a-Sedano, Garofalo, Garralda, Gavras, Gerssen, Geyer, Gilmore,
  Girona, Giuffrida, Gomes, Gonz{\'a}lez-Marcos, Gonz{\'a}lez-N{\'u}{\~n}ez,
  Gonz{\'a}lez-Vidal, Granvik, Guerrier, Guillout, Guiraud, G{\'u}rpide,
  Guti{\'e}rrez-S{\'a}nchez, Guy, Haigron, Hatzidimitriou, Haywood, Heiter,
  Helmi, Hobbs, Hofmann, Holl, Holland, Hunt, Hypki, Icardi, Irwin, Jevardat~de
  Fombelle, Jofre, Jonker, Jorissen, Julbe, Karampelas, Kochoska, Kohley,
  Kolenberg, and Kontizas}]{GaiaCollaboration_2016}
{Gaia Collaboration}, Brown AGA, Vallenari A et~al. (2016) {Gaia Data Release
  1}. Astronomy and Astrophysics 595:A2

\bibitem[{{Gillon} et~al.(2017){Gillon}, {Triaud}, {Demory}, {Jehin}, {Agol},
  {Deck}, {Lederer}, {de Wit}, {Burdanov}, {Ingalls}, {Bolmont}, {Leconte},
  {Raymond}, {Selsis}, {Turbet}, {Barkaoui}, {Burgasser}, {Burleigh}, {Carey},
  {Chaushev}, {Copperwheat}, {Delrez}, {Fernandes}, {Holdsworth}, {Kotze}, {Van
  Grootel}, {Almleaky}, {Benkhaldoun}, {Magain}, and {Queloz}}]{gillon_etal_17}
{Gillon} M, {Triaud} AHMJ, {Demory} BO et~al. (2017) {Seven temperate
  terrestrial planets around the nearby ultracool dwarf star TRAPPIST-1}. \nat
  542:456--460

\bibitem[{{Hanbury Brown} et~al.(1974){Hanbury Brown}, {Davis}, and
  {Allen}}]{Hanbury_Brown_74}
{Hanbury Brown} R, {Davis} J {Allen} LR (1974) {The Angular Diameters of 32
  Stars}. \mnras 167:121--136

\bibitem[{Huber et~al.(2012)Huber, Ireland, Bedding, Brand{\~a}o, Piau,
  Maestro, White, Bruntt, Casagrande, Molenda-{\.Z}akowicz, Aguirre, Sousa,
  Barclay, Burke, Chaplin, Christensen-Dalsgaard, Cunha, De~Ridder, Farrington,
  Frasca, Garc{\'\i}a, Gilliland, Goldfinger, Hekker, Kawaler, Kjeldsen,
  Mcalister, Metcalfe, Miglio, Monteiro, Pinsonneault, Schaefer, Stello,
  Stumpe, Sturmann, Sturmann, Ten~Brummelaar, Thompson, Turner, and
  Uytterhoeven}]{Huber_2012}
Huber D, Ireland MJ, Bedding TR et~al. (2012) {Fundamental properties of stars
  using asteroseismology from Kepler and CoRoT and interferometry from the
  CHARA Array}. ApJ 760(1):32

\bibitem[{{Jeffries}(2014)}]{jeffries_14}
{Jeffries} RD (2014) {Using rotation, magnetic activity and lithium to estimate
  the ages of low mass stars}. In: EAS Publications Series, EAS Publications
  Series, vol~65, pp 289--325, \doi{10.1051/eas/1465008}

\bibitem[{{J{\o}rgensen} and {Lindegren}(2005)}]{jorgensen_lindegren_05}
{J{\o}rgensen} BR {Lindegren} L (2005) {Determination of stellar ages from
  isochrones: Bayesian estimation versus isochrone fitting}. \aap 436:127--143

\bibitem[{{Kippenhahn} et~al.(2012){Kippenhahn}, {Weigert}, and
  {Weiss}}]{kippenhahn_etal_12}
{Kippenhahn} R, {Weigert} A {Weiss} A (2012) {Stellar Structure and Evolution}.
  \doi{10.1007/978-3-642-30304-3}

\bibitem[{{Mamajek} and {Hillenbrand}(2008)}]{mamajek_hillenbrand_07}
{Mamajek} EE {Hillenbrand} LA (2008) {Improved Age Estimation for Solar-Type
  Dwarfs Using Activity-Rotation Diagnostics}. \apj 687:1264-1293

\bibitem[{{Mann} et~al.(2016{\natexlab{a}}){Mann}, {Gaidos}, {Mace}, {Johnson},
  {Bowler}, {LaCourse}, {Jacobs}, {Vanderburg}, {Kraus}, {Kaplan}, and
  {Jaffe}}]{mann_etal_16a}
{Mann} AW, {Gaidos} E, {Mace} GN et~al. (2016{\natexlab{a}}) {Zodiacal
  Exoplanets in Time (ZEIT). I. A Neptune-sized Planet Orbiting an M4.5 Dwarf
  in the Hyades Star Cluster}. \apj 818:46

\bibitem[{{Mann} et~al.(2016{\natexlab{b}}){Mann}, {Newton}, {Rizzuto},
  {Irwin}, {Feiden}, {Gaidos}, {Mace}, {Kraus}, {James}, {Ansdell},
  {Charbonneau}, {Covey}, {Ireland}, {Jaffe}, {Johnson}, {Kidder}, and
  {Vanderburg}}]{mann_etal_16b}
{Mann} AW, {Newton} ER, {Rizzuto} AC et~al. (2016{\natexlab{b}}) {Zodiacal
  Exoplanets in Time (ZEIT). III. A Short-period Planet Orbiting a
  Pre-main-sequence Star in the Upper Scorpius OB Association}. \aj 152:61

\bibitem[{{Mann} et~al.(2017){Mann}, {Gaidos}, {Vanderburg}, {Rizzuto},
  {Ansdell}, {Medina}, {Mace}, {Kraus}, and {Sokal}}]{mann_etal_17a}
{Mann} AW, {Gaidos} E, {Vanderburg} A et~al. (2017) {Zodiacal Exoplanets in
  Time (ZEIT). IV. Seven Transiting Planets in the Praesepe Cluster}. \aj
  153:64

\bibitem[{{Mann} et~al.(2018){Mann}, {Vanderburg}, {Rizzuto}, {Kraus},
  {Berlind}, {Bieryla}, {Calkins}, {Esquerdo}, {Latham}, {Mace}, {Morris},
  {Quinn}, {Sokal}, and {Stefanik}}]{mann_etal_18}
{Mann} AW, {Vanderburg} A, {Rizzuto} AC et~al. (2018) {Zodiacal Exoplanets in
  Time (ZEIT). VI. A Three-planet System in the Hyades Cluster Including an
  Earth-sized Planet}. \aj 155:4

\bibitem[{{Maxted} et~al.(2015{\natexlab{a}}){Maxted}, {Serenelli}, and
  {Southworth}}]{maxted_etal_15a}
{Maxted} PFL, {Serenelli} AM {Southworth} J (2015{\natexlab{a}}) {Bayesian mass
  and age estimates for transiting exoplanet host stars}. \aap 575:A36

\bibitem[{{Maxted} et~al.(2015{\natexlab{b}}){Maxted}, {Serenelli}, and
  {Southworth}}]{maxted_etal_15b}
{Maxted} PFL, {Serenelli} AM {Southworth} J (2015{\natexlab{b}}) {Comparison of
  gyrochronological and isochronal age estimates for transiting exoplanet host
  stars}. \aap 577:A90

\bibitem[{{Meibom} et~al.(2009){Meibom}, {Mathieu}, and
  {Stassun}}]{meibom_etal_09}
{Meibom} S, {Mathieu} RD {Stassun} KG (2009) {Stellar Rotation in M35:
  Mass-Period Relations, Spin-Down Rates, and Gyrochronology}. \apj
  695:679--694

\bibitem[{{Meibom} et~al.(2011{\natexlab{a}}){Meibom}, {Barnes}, {Latham},
  {Batalha}, {Borucki}, {Koch}, {Basri}, {Walkowicz}, {Janes}, {Jenkins}, {Van
  Cleve}, {Haas}, {Bryson}, {Dupree}, {Furesz}, {Szentgyorgyi}, {Buchhave},
  {Clarke}, {Twicken}, and {Quintana}}]{meibom_etal_11}
{Meibom} S, {Barnes} SA, {Latham} DW et~al. (2011{\natexlab{a}}) {The Kepler
  Cluster Study: Stellar Rotation in NGC 6811}. \apjl 733:L9

\bibitem[{{Meibom} et~al.(2011{\natexlab{b}}){Meibom}, {Mathieu}, {Stassun},
  {Liebesny}, and {Saar}}]{meibom_etal_11b}
{Meibom} S, {Mathieu} RD, {Stassun} KG, {Liebesny} P {Saar} SH
  (2011{\natexlab{b}}) {The Color-period Diagram and Stellar Rotational
  Evolution{---}New Rotation Period Measurements in the Open Cluster M34}. \apj
  733:115

\bibitem[{{Meibom} et~al.(2015){Meibom}, {Barnes}, {Platais}, {Gilliland},
  {Latham}, and {Mathieu}}]{meibom_etal_15}
{Meibom} S, {Barnes} SA, {Platais} I et~al. (2015) {A spin-down clock for cool
  stars from observations of a 2.5-billion-year-old cluster}. \nat 517:589--591

\bibitem[{{Pont} and {Eyer}(2004)}]{pont_eyer_04}
{Pont} F {Eyer} L (2004) {Isochrone ages for field dwarfs: method and
  application to the age-metallicity relation}. \mnras 351:487--504

\bibitem[{{Rauer} et~al.(2014){Rauer}, {Catala}, {Aerts}, {Appourchaux},
  {Benz}, {Brandeker}, {Christensen-Dalsgaard}, {Deleuil}, {Gizon}, {Goupil},
  {G{\"u}del}, {Janot-Pacheco}, {Mas-Hesse}, {Pagano}, {Piotto}, {Pollacco},
  {Santos}, {Smith}, {Su{\'a}rez}, {Szab{\'o}}, {Udry}, {Adibekyan}, {Alibert},
  {Almenara}, {Amaro-Seoane}, {Eiff}, {Asplund}, {Antonello}, {Barnes},
  {Baudin}, {Belkacem}, {Bergemann}, {Bihain}, {Birch}, {Bonfils}, {Boisse},
  {Bonomo}, {Borsa}, {Brand{\~a}o}, {Brocato}, {Brun}, {Burleigh}, {Burston},
  {Cabrera}, {Cassisi}, {Chaplin}, {Charpinet}, {Chiappini}, {Church},
  {Csizmadia}, {Cunha}, {Damasso}, {Davies}, {Deeg}, {D{\'{\i}}az}, {Dreizler},
  {Dreyer}, {Eggenberger}, {Ehrenreich}, {Eigm{\"u}ller}, {Erikson}, {Farmer},
  {Feltzing}, {de Oliveira Fialho}, {Figueira}, {Forveille}, {Fridlund},
  {Garc{\'{\i}}a}, {Giommi}, {Giuffrida}, {Godolt}, {Gomes da Silva},
  {Granzer}, {Grenfell}, {Grotsch-Noels}, {G{\"u}nther}, {Haswell}, {Hatzes},
  {H{\'e}brard}, {Hekker}, {Helled}, {Heng}, {Jenkins}, {Johansen},
  {Khodachenko}, {Kislyakova}, {Kley}, {Kolb}, {Krivova}, {Kupka}, {Lammer},
  {Lanza}, {Lebreton}, {Magrin}, {Marcos-Arenal}, {Marrese}, {Marques},
  {Martins}, {Mathis}, {Mathur}, {Messina}, {Miglio}, {Montalban}, {Montalto},
  {Monteiro}, {Moradi}, {Moravveji}, {Mordasini}, {Morel}, {Mortier},
  {Nascimbeni}, {Nelson}, {Nielsen}, {Noack}, {Norton}, {Ofir}, {Oshagh},
  {Ouazzani}, {P{\'a}pics}, {Parro}, {Petit}, {Plez}, {Poretti}, {Quirrenbach},
  {Ragazzoni}, {Raimondo}, {Rainer}, {Reese}, {Redmer}, {Reffert},
  {Rojas-Ayala}, {Roxburgh}, {Salmon}, {Santerne}, {Schneider}, {Schou},
  {Schuh}, {Schunker}, {Silva-Valio}, {Silvotti}, {Skillen}, {Snellen}, {Sohl},
  {Sousa}, {Sozzetti}, {Stello}, {Strassmeier}, {{\v S}vanda}, {Szab{\'o}},
  {Tkachenko}, {Valencia}, {Van Grootel}, {Vauclair}, {Ventura}, {Wagner},
  {Walton}, {Weingrill}, {Werner}, {Wheatley}, and {Zwintz}}]{rauer_etal_14}
{Rauer} H, {Catala} C, {Aerts} C et~al. (2014) {The PLATO 2.0 mission}.
  Experimental Astronomy 38:249--330

\bibitem[{{Ricker} et~al.(2016){Ricker}, {Vanderspek}, {Winn}, {Seager},
  {Berta-Thompson}, {Levine}, {Villasenor}, {Latham}, {Charbonneau}, {Holman},
  {Johnson}, {Sasselov}, {Szentgyorgyi}, {Torres}, {Bakos}, {Brown},
  {Christensen-Dalsgaard}, {Kjeldsen}, {Clampin}, {Rinehart}, {Deming}, {Doty},
  {Dunham}, {Ida}, {Kawai}, {Sato}, {Jenkins}, {Lissauer}, {Jernigan},
  {Kaltenegger}, {Laughlin}, {Lin}, {McCullough}, {Narita}, {Pepper},
  {Stassun}, and {Udry}}]{ricker_etal_16}
{Ricker} GR, {Vanderspek} R, {Winn} J et~al. (2016) {The Transiting Exoplanet
  Survey Satellite}. In: Space Telescopes and Instrumentation 2016: Optical,
  Infrared, and Millimeter Wave, \procspie, vol 9904, p 99042B,
  \doi{10.1117/12.2232071}

\bibitem[{{Sandford} and {Kipping}(2017)}]{sandford_17}
{Sandford} E {Kipping} D (2017) {Know the Planet, Know the Star: Precise
  Stellar Densities from Kepler Transit Light Curves}. \aj 154:228

\bibitem[{Segransan et~al.(2003)Segransan, Kervella, Forveille, and
  Queloz}]{Segransan_2003}
Segransan D, Kervella P, Forveille T Queloz D (2003) {First radius measurements
  of very low mass stars with the VLTI}. Astronomy and Astrophysics
  397(3):L5--L8

\bibitem[{{Serenelli} et~al.(2013){Serenelli}, {Bergemann}, {Ruchti}, and
  {Casagrande}}]{serenelli_etal_13}
{Serenelli} AM, {Bergemann} M, {Ruchti} G {Casagrande} L (2013) {Bayesian
  analysis of ages, masses and distances to cool stars with non-LTE
  spectroscopic parameters}. \mnras 429:3645--3657

\bibitem[{{Sestito} and {Randich}(2005)}]{sestito_randich_05}
{Sestito} P {Randich} S (2005) {Time scales of Li evolution: a homogeneous
  analysis of open clusters from ZAMS to late-MS}. \aap 442:615--627

\bibitem[{{Silva Aguirre} et~al.(2013){Silva Aguirre}, {Basu}, {Brand{\~a}o},
  {Christensen-Dalsgaard}, {Deheuvels}, {Do{\u g}an}, {Metcalfe}, {Serenelli},
  {Ballot}, {Chaplin}, {Cunha}, {Weiss}, {Appourchaux}, {Casagrande},
  {Cassisi}, {Creevey}, {Garc{\'{\i}}a}, {Lebreton}, {Noels}, {Sousa},
  {Stello}, {White}, {Kawaler}, and {Kjeldsen}}]{silva_aguirre_etal_13}
{Silva Aguirre} V, {Basu} S, {Brand{\~a}o} IM et~al. (2013) {Stellar Ages and
  Convective Cores in Field Main-sequence Stars: First Asteroseismic
  Application to Two Kepler Targets}. \apj 769:141

\bibitem[{{Silva Aguirre} et~al.(2015){Silva Aguirre}, {Davies}, {Basu},
  {Christensen-Dalsgaard}, {Creevey}, {Metcalfe}, {Bedding}, {Casagrande},
  {Handberg}, {Lund}, {Nissen}, {Chaplin}, {Huber}, {Serenelli}, {Stello}, {Van
  Eylen}, {Campante}, {Elsworth}, {Gilliland}, {Hekker}, {Karoff}, {Kawaler},
  {Kjeldsen}, and {Lundkvist}}]{silva_aguirre_etal_15}
{Silva Aguirre} V, {Davies} GR, {Basu} S et~al. (2015) {Ages and fundamental
  properties of Kepler exoplanet host stars from asteroseismology}. \mnras
  452:2127--2148

\bibitem[{{Silva Aguirre} et~al.(2017){Silva Aguirre}, {Lund}, {Antia}, {Ball},
  {Basu}, {Christensen-Dalsgaard}, {Lebreton}, {Reese}, {Verma}, {Casagrande},
  {Justesen}, {Mosumgaard}, {Chaplin}, {Bedding}, {Davies}, {Handberg},
  {Houdek}, {Huber}, {Kjeldsen}, {Latham}, {White}, {Coelho}, {Miglio}, and
  {Rendle}}]{silva_aguirre_etal_17}
{Silva Aguirre} V, {Lund} MN, {Antia} HM et~al. (2017) {Standing on the
  Shoulders of Dwarfs: the Kepler Asteroseismic LEGACY Sample. II.Radii,
  Masses, and Ages}. \apj 835:173

\bibitem[{{Skumanich}(1972)}]{skumanich_72}
{Skumanich} A (1972) {Time Scales for CA II Emission Decay, Rotational Braking,
  and Lithium Depletion}. \apj 171:565

\bibitem[{{Soderblom}(2010)}]{soderblom_10}
{Soderblom} DR (2010) {The Ages of Stars}. \araa 48:581--629

\bibitem[{{Soderblom} et~al.(2014){Soderblom}, {Hillenbrand}, {Jeffries},
  {Mamajek}, and {Naylor}}]{soderblom_etal_14}
{Soderblom} DR, {Hillenbrand} LA, {Jeffries} RD, {Mamajek} EE {Naylor} T (2014)
  {Ages of Young Stars}. Protostars and Planets VI pp 219--241

\bibitem[{{Somers} and {Pinsonneault}(2016)}]{somers_pinsonneault_16}
{Somers} G {Pinsonneault} MH (2016) {Lithium Depletion is a Strong Test of
  Core-envelope Recoupling}. \apj 829:32

\bibitem[{{Thompson} et~al.(2003){Thompson}, {Christensen-Dalsgaard}, {Miesch},
  and {Toomre}}]{thompson_etal_03}
{Thompson} MJ, {Christensen-Dalsgaard} J, {Miesch} MS {Toomre} J (2003) {The
  Internal Rotation of the Sun}. \araa 41:599--643

\bibitem[{{Triaud}(2011)}]{triaud_11}
{Triaud} AHMJ (2011) {The time dependence of hot Jupiters' orbital
  inclinations}. \aap 534:L6

\bibitem[{{van Saders} et~al.(2016){van Saders}, {Ceillier}, {Metcalfe}, {Silva
  Aguirre}, {Pinsonneault}, {Garc{\'{\i}}a}, {Mathur}, and
  {Davies}}]{van_saders_etal_16}
{van Saders} JL, {Ceillier} T, {Metcalfe} TS et~al. (2016) {Weakened magnetic
  braking as the origin of anomalously rapid rotation in old field stars}. \nat
  529:181--184

\bibitem[{White et~al.(2013)White, Huber, Maestro, Bedding, Ireland, Baron,
  Boyajian, Che, Monnier, Pope, Roettenbacher, Stello, Tuthill, Farrington,
  Goldfinger, Mcalister, Schaefer, Sturmann, Sturmann, Ten~Brummelaar, and
  Turner}]{White_2013}
White TR, Huber D, Maestro V et~al. (2013) {Interferometric radii of bright
  Kepler stars with the CHARA Array: Cygni and 16 Cygni A and B}. MNRAS
  433(2):1262--1270

\bibitem[{{Zwintz} et~al.(2014){Zwintz}, {Fossati}, {Ryabchikova}, {Guenther},
  {Aerts}, {Barnes}, {Theme{\ss}l}, {Lorenz}, {Cameron}, {Kuschnig},
  {Pollack-Drs}, {Moravveji}, {Baglin}, {Matthews}, {Moffat}, {Poretti},
  {Rainer}, {Rucinski}, {Sasselov}, and {Weiss}}]{zwintz_etal_14}
{Zwintz} K, {Fossati} L, {Ryabchikova} T et~al. (2014) {Echography of young
  stars reveals their evolution}. Science 345:550--553

\end{thebibliography}

%
\end{document}